# Geometric phases for a thermal two-dimensional mixed spin 1/2 system

Y. Ben-Aryeh

Physics Department, Technion-Israel Institute of Technology, Haifa, 32000. Israel

E-mail: phr65yb@physics.technion.ac.il

**Abstract**

Quantum mechanical methods for getting geometric phases for mixed states are analyzed. Parallel transport equations for pure states are generalized to mixed states by which dynamical phases are eliminated. The geometric phases are derived by SU(2) transformations of mixed states which are different from those used in NMR and neutron interference experiments. The mixed states are obtained as two-dimensional thermal states. The SU(2) transformations include time dependent parameters which are function of the magnetic fields interactions in times which are short relative to relaxation times. These conditions lead to a special form of the unitary transformation of the mixed thermal states by which results for geometric phase and for interference intensity are derived.

## 1. Introduction

Derivations of geometric phases are based on topological effects [1,2] which can be analyzed either classically or quantum mechanically. For example, the dynamics of charge pumping in a dot coupled to two reservoirs were analyzed for getting geometric phases by using classical methods based on rate equations including the fluctuations [3]. Quantum mechanical methods for getting geometric phases were developed mainly for pure states. In one of the first papers on this topic it was shown by Berry that within the adiabatic proximation a pure quantal state which undergoes cyclic evolution may acquire geometric phase in addition to the dynamical phase [4]. Later this effect was generalized to the non-adiabatic case [5]. Related to Pancharatnam's earlier work [6] geometric phases were developed to non-cyclic evolution of quantum systems [7,8]. Enormous number of applications of geometric phases for different pure quantal systems were developed.

The topic of geometric phases for mixed states in interferometry was developed by using unitary transformation of mixed states [9]. Following this work we consider a mixed state in N-dimensional

Hilbert space where its initial density matrix $\rho_0$ is diagonal and spanned by the ortho-normal states $|k\rangle, k = 1, 2, ..., N$ as given at time $t = 0$ by:

$$\rho_0 = \sum_k w_k |k\rangle_0 \langle k|_0 \tag{1}$$

where the short notation $|k\rangle_0$ denotes individual pure states at time $t = 0$ with classical probability $w_k$. For simplicity we assume that initial vectors $|k\rangle_0$ are real and represented, for example, for $N = 3$ as

$$|1\rangle_0 = \begin{pmatrix} 1 \\ 0 \\ 0 \end{pmatrix} \; ; \; |2\rangle_0 = \begin{pmatrix} 0 \\ 1 \\ 0 \end{pmatrix} \; ; \; |3\rangle_0 = \begin{pmatrix} 0 \\ 0 \\ 1 \end{pmatrix} \tag{2}$$

and similarly for other $N$ values.

We study here the possibility to obtain geometric phases in thermal states. We consider zeroth order Hamiltonian $\hat{H}_0$ of interaction between a magnetic moment $\vec{M}$ and a magnetic field $\vec{B}$ given by $\hat{H}_0 = -\vec{M} \cdot \vec{B}$. We assume a constant magnetic field $B_z$ which is along the positive $z$ axis. We specialize our example to the case of spin 1/2 system where the Hamiltonian $\hat{H}_0$ is:

$$\hat{H}_0 = \hbar \tilde{\omega} \hat{s}_z \; ; \; \tilde{\omega} = 2\mu_B B_z \tag{3}$$

where $\mu_B$ is the Bohr magneton. The mixed density matrix $\rho_0$ of this system is given by:

$$\rho_0 = w_1 |-1/2\rangle \langle -1/2| + w_2 |+1/2\rangle \langle +1/2| \tag{4}$$

where $\{-1/2, +1/2\}$ are the eigenvalues of the spin and the classical probabilities $w_1$ and $w_2$ are given respectively, by statistical mechanics as:

$$w_1 = \frac{e^{\hbar\tilde{\omega}\beta/2}}{e^{\hbar\tilde{\omega}\beta/2} + e^{-\hbar\tilde{\omega}\beta/2}} \; ; \; w_2 = \frac{e^{-\hbar\tilde{\omega}\beta/2}}{e^{\hbar\tilde{\omega}\beta/2} + e^{-\hbar\tilde{\omega}\beta/2}} \; ; \; \beta = \frac{1}{k_B T} \tag{5}$$

where $k_B$ is the Boltzmann constant, $T$ the absolute temperature, $\mu_B$ is the Bohr magneton and $\tilde{\omega}$ was defined in Eq. (3). We refer to $\tilde{\omega}$ rather than $B_z$ as "the field". Based on statistical mechanics, the expectation value of $\hat{S}_z$ is given by

$$\langle s_z \rangle = -\frac{1}{2}\tanh(\tilde{\omega}\beta/2) \quad ; \quad \beta = \frac{1}{k_b T} \tag{6}$$

The internal energy $E_0$ of the spin system is given by the expectation value of the Hamiltonian $\hat{H}_0$:

$$E_0 = \langle H_0 \rangle = \hbar\tilde{\omega}\langle s_z \rangle \tag{7}$$

Quantum thermal engines connected to a hot and cold heat bath have been studied extensively in various forms [3,10-12]. Such machines work by different steps in which for example, the energy of spin 1/2 step (considered as the working medium) is given by the expectation value of Eq. (7). In each step the temperature or the "frequency" $\tilde{\omega} = 2\mu_B B_z$ (or both parameters) are changing. The efficiency of such machines is calculated under different conditions including various perturbations. The purpose of the present analysis is different as we consider the terms proportional to $w_1$ and $w_2$ in Eq. (5) to be the components of mixed state and study the possibility to get geometric phases by performing a special kind of unitary transformation operating separately on these components.

We present in the next Section the general theory by which geometric phases are obtained for mixed states and dynamical phases are eliminated by using parallel transport conditions. The main results are presented in Section 3 where we analyze the method by which geometric phases can be obtained by unitary transformation of thermal states. Thermal states are used as an important component in thermal engines where their time development involves usually only dynamical phases [3]. I hope that the possibility to get geometric phases in quantum thermal engines will be of interest.

## 2. General theory for geometric phases obtained in mixed states

The density matrix $\rho_0$ for mixed states given in Eq. (1) includes averaging of thermal fluctuations described by the probabilities $w_k$. Therefore, unitary transformations cannot operate directly on this density matrix, but they can operate separately on the components of such density matrix composed of pure quantum states. The idea presented in [9] is that the initial density matrix $\rho_0$ can be developed in time to a density matrix $\rho_t$ by the unitary transformation operating on the vectors $|k\rangle_0$ as

$$\rho_t = \sum_{k=1}^{N} w_k \left\{ U(t)|k\rangle_0 \langle k|_0 U^\dagger(t) \right\} = \sum_{k=1}^{N} w_k |k\rangle_t \langle k|_t \qquad (8)$$

where $U(t)$ is a unitary matrix of $N \times N$ dimension, $|k\rangle_t = U(t)|k\rangle_0$ are the orthonormal vectors at time $t$ and $|k\rangle_0$ are the initial vectors. The mixed state is described as a mixture of several pure states incoherently weighed by their respective probabilities $w_k$. For thermal states the eigenvectors $|1\rangle_0 = \begin{pmatrix} 1 \\ 0 \end{pmatrix}$ and $|2\rangle_0 = \begin{pmatrix} 0 \\ 1 \end{pmatrix}$ correspond to eigenvalues -1/2 and +1/2 of $\hat{s}_z$ and the unitary matrix is two-dimensional.

The parallel transport of a particular vector $|k\rangle_t$ implies no change in phase when $|k\rangle_t$ evolves into $|k\rangle_{t+dt}$, for some infinitesimal change of the parameter $t$. Although locally there is no phase change, the system can acquire a geometric phase after completing a closed loop parametrized by $t$. This phase is related to curvature of the parameter space depending only on the geometry of the path. The parallel transport can be satisfied by choosing a suitable $U(t)$ matrix which satisfies the parallel transport conditions:

$$\langle k|_t \dot{U}(t) U^\dagger(t) |k\rangle_t = 0 \quad ; \quad k = 1, 2, \cdots, N \qquad (9)$$

Eq. (9) represents generalization of the parallel transport equation $Tr\left[\rho_t \dot{U}(t) U^\dagger(t)\right] = 0$ for pure states to that of mixed states [9].

The geometric phase $\gamma_{geom}$ acquired by a mixed state evolving along a closed curve $\Gamma$ under unitary transformation which satisfy the parallel transport Eqs. (9) is given by:

$$\gamma_{geom}(\Gamma) = \arg Tr\left[\rho_t U(t)\right] = \arg Tr\left[U(t) \rho_0 U^\dagger(t) U(t)\right] = \arg Tr\left[U(t) \rho_0\right] \quad . \qquad (10)$$

By substituting in Eq. (10) the density matrix $\rho_0$ according to Eq. (1) and assuming the parallel transport conditions we get by straightforward calculations:

$$\langle k|_0 U(t) |k\rangle_0 = v_k e^{i\beta_k} \quad ; \quad \gamma_{geom} = \arg\left[\sum_k w_k v_k e^{i\beta_k}\right] \qquad (11)$$

where $v_k$ and $\beta_k$ are, respectively, the absolute value and phase of $\langle k|_0 U(t)|k\rangle_0$.

By straightforward calculations we can transform the sum of Eq. (11) into the form

$$\sum_k w_k v_k e^{i\beta_k} = \tilde{v} e^{i\theta} \quad ; \quad \gamma_{geom} = \theta \tag{12}$$

where $\tilde{v}$ and $\theta$ are, respectively, the absolute value and the geometric phase given for the sum on the right side of Eq. (11) and where $\tilde{v}$ represents the interference intensity.

Quantal geometric phase in split-beam interferometry was developed by using non-cyclic SU(2) spatial evolution which is given as Pancharatnam's phase [13]. Noncyclic Pancharatnam's phase for mixed state SU(2) evolution in neutron interferometry was developed in [14]. We would like to use unitary transformation of the mixed states of Eq. (1) by following the analysis of Eqs. (9-12) which leads to geometric phase. The phase $\beta_k$ of Eq. (11) represents Pancharatnam's phase difference between $|k\rangle_0$ and $|k\rangle_t$ where these phase differences are gauge invariant [2] so that the above equations can be used also for non-cyclic evolutions. The phase $\theta$ represents the geometric Pancharatnam's phase where parameters $v_k < 1$ and $\tilde{v} < 1$ represent reduced intensities in the interference experiments.

Measurements of geometric phases for mixed states were made for NMR experiments [15-17] in which the density operator of spin 1/2 particle can be written as

$$\rho_t = \frac{1}{2}(1 + \vec{r} \cdot \vec{\sigma}) \quad . \tag{13}$$

Here $r$ is the length of the Bloch vector which remains unchanged during evolution of the state ($r$ less than 1 represents mixed states). The components of $\vec{\sigma}$ are the Pauli matrices and $\rho_t$ represents a mixture of its eigenvectors with eigenvalues $\frac{1}{2}(1 \pm r)$. A Bloch vector for mixed state ($r < 1$) traces out a cyclic curve in the Bloch sphere and which acquires a geodesic closed solid angle $\Omega$. The two eigenvalues, acquire respectively, phase $\pm \Omega/2$ and in this case $\gamma_{geom}(\Gamma)$ acquires the geometric phase:

$$\gamma_{geom}(\Gamma) = \frac{1}{2}(1-r)e^{i\Omega/2} + \frac{1}{2}(1+r)e^{-i\Omega/2} = \cos(\Omega/2) - ir\sin(\Omega/2) \quad . \tag{14}$$

Eq. (14) can be transformed into the form of Eq, (12) where:

$$\tilde{v} = \sqrt{\cos^2(\Omega/2) + r^2 \sin^2(\Omega/2)} \quad ; \quad \gamma_{geom} = -\arctan(r\tan(\Omega/2)) \tag{15} \quad .$$

It is interesting to note that time development of quantum engines with the working medium of spin 1/2 system depends on two parameters: magnetic field and temperature. In the analysis presented in the next Section for mixed states temperature enters in the parameters $w_1$ and $w_2$ obtained in Eq. (5) by thermal averaging, while the dynamics enters by the Hamiltonians including magnetic fields interactions. Usually, one starts with unitary operators and then add noise terms. We use a different approach in the next Section by which we start with mixed state which has already noise effects and only afterwards add the unitary operators. Such new approach may have general implications for treating quantum noise but here we concentrate on the idea that the above unitary transformation of mixed state can replace the isothermal steps in quantum engines. Quite common approach to add noise to spin 1/2 system is to change $\vec{\sigma}$ into the product $\vec{\sigma} \cdot \vec{r}$ for the density matrix like that used in Eq. (13) for NMR and for neutron interference experiments in [14] (where $|\vec{r}| < 1$ represent the noise effect). While such approach can help in explaining certain experiments the inclusion of thermal effects are not obvious. It seems that the new approach described in the next section of using unitary transformation of mixed states and inserting the thermal noise effects in the initial state are more suitable for describing isothermal quantum steps in quantum engines which can include geometric phases.

Relations between angular momentum operators and unitary transformations were developed in many works. Espeially relations between such unitary transformations and dynamical and geometric phases were treated [18]. Also, geometric phases for N-level systems were treated by using SU(N) transformations [19]. Although these works are interesting the present topic is different as we study the geometric phases obtained from unitary transformations of mixed states. In the next Section we study the possibility that the above methods can be used to get geometric phases in thermal states.

## 3. Geometric phases for thermal states in interferometry

Geometric phases for thermal states can be obtained by using the transformations:

$$|1\rangle_t = U(t) \begin{pmatrix} 1 \\ 0 \end{pmatrix}_0 \quad ; \quad |2\rangle_2 = U(t) \begin{pmatrix} 0 \\ 1 \end{pmatrix}_0 \qquad . \qquad (16)$$

where $U(t)$ is the SU(2), 2-dimensional unitary matrix which satisfy the parallel transport equations and is decomposed as:

$$SU(2) = U(t) = U_1(\alpha)U_2(\beta)U_3(\gamma) \tag{17}$$

Here:

$$U_1(\alpha) = \begin{bmatrix} \exp[i\alpha/2 + \phi] & 0 \\ 0 & \exp[-i\alpha/2 + \phi] \end{bmatrix} ; \quad U_3(\gamma) = \begin{bmatrix} \exp[i\gamma/2] & 0 \\ 0 & \exp[-i\gamma/2] \end{bmatrix}$$

$$U_2(\beta) = \begin{bmatrix} \cos(\beta/2) & \sin(\beta/2) \\ -\sin(\beta/2) & \cos(\beta/2) \end{bmatrix} \tag{18}$$

where $\phi$ is a certain phase shift between $U_1(\alpha)$ and $U_3(\gamma)$. We assume that $U_3(\gamma)$ and $U_1(\gamma)$ operate during time $t$ with Hamiltonians fixed by Eq, (3) as $H_{1z} = -\hbar\tilde{\omega}_1\sigma_z/2$ and $H_{2z} = -\hbar\tilde{\omega}_2\sigma_z/2$ respectively, where $\tilde{\omega}_1 = 2\mu_B B_{1z}$ and $\tilde{\omega}_2 = 2\mu B_{2z}$, and where $B_{1z}$ and $B_{2z}$ are constants magnetic fields in the $z$ direction. Then we get:

$$U_1(\alpha) = \begin{bmatrix} \exp[i\tilde{\omega}_1 t/2 + \phi] & 0 \\ 0 & \exp[-i\tilde{\omega}_1 t/2 + \phi] \end{bmatrix} ; \quad U_3(\gamma) = \begin{bmatrix} \exp[i\tilde{\omega}_2 t/2] & 0 \\ 0 & \exp[-i\tilde{\omega}_2 t/2] \end{bmatrix} \tag{19}$$

$U_2(\beta)$ is a simple beam-splitter transformation which does not include any time dependence and for simplicity is defined as:

$$U_2(\beta) = \begin{bmatrix} \cos\xi & \sin\xi \\ -\sin\xi & \cos\xi \end{bmatrix} ; \quad \xi = \frac{\beta}{2} \tag{20}$$

By substituting Eqs. (19-20) into Eqs. (17-18) we get:

$$U(t) = U_1(t)U_2(t)U_3(t) = e^{i\phi}\begin{pmatrix} e^{i\delta t}\cos\xi & -e^{-i\varsigma t}\sin\xi \\ e^{i\varsigma t}\sin\xi & e^{-i\delta t}\cos\xi \end{pmatrix} \tag{21}$$

where

$$\delta = \frac{\tilde{\omega}_1 + \tilde{\omega}_2}{2} ; \quad \varsigma = \frac{\tilde{\omega}_2 - \tilde{\omega}_1}{2} \tag{22}$$

Eq. (21) is similar to the SU(2) transformation used in in neutron interferometry ([14] Eq.(1)). But here the parameters $\delta$ and $\varsigma$ are multiplied by the time $t$ which leads to a certain dependence of $U$ on time which is critical for the use of the following analysis.

Realization of the above unitary transformation depends on scales of time. The initial mixed state of Eqs. (4-5) can be produced after relatively long time in which the SU(2) system arrives at thermal equilibrium. The operation of the unitary transformation on the initial state can be made during time t which is short relative to relaxation times.

We would like to use the above $U(t)$ unitary matrix for obtaining parallel transport equations. Due to the linear relations between $|k\rangle_t$ and $|k\rangle_0$ the two parallel transport equations obtained from Eq, (9) as $\langle k|_t \dot{U}(t) U^\dagger(t) |k\rangle_t = 0$ ; $k = 1, 2$ implies also:

$$\langle k|_0 \dot{U}(t) U^\dagger(t) |k\rangle_0 = 0 \; ; \; k = 1, 2 \tag{23}$$

For the first equation we get:

$$(1, 0) \begin{pmatrix} (i\delta \cos\xi) e^{i\delta t} & (i\varsigma \sin\xi) e^{-i\varsigma t} \\ (i\varsigma \sin\xi) e^{i\varsigma t} & (-i\delta \cos\xi) e^{-i\delta t} \end{pmatrix}$$
$$\times \begin{pmatrix} \cos\xi e^{-i\delta t} & \sin\xi e^{-i\varsigma t} \\ -\sin\xi e^{i\varsigma t} & \cos\xi e^{i\delta t} \end{pmatrix} \begin{pmatrix} 1 \\ 0 \end{pmatrix} = 0 \tag{24}$$

Straightforward calculations lead to the result:

$$\left( (i\delta \cos\xi) e^{i\delta t} \quad (i\varsigma \sin\xi) e^{-i\varsigma t} \right)$$
$$\times \begin{pmatrix} \cos\xi e^{-i\delta t} \\ -\sin\xi e^{i\varsigma t} \end{pmatrix} = -i\delta (\cos\xi)^2 + i\varsigma (\sin\xi)^2 = 0 \tag{25}$$

For the second equation we use calculations from Eq. (24) in which we exchange $(1 \; 0)$ to $(0 \; 1)$ and $\begin{pmatrix} 1 \\ 0 \end{pmatrix}$ to $\begin{pmatrix} 0 \\ 1 \end{pmatrix}$. Then we get instead of Eq. (25):

$$\left( i\varsigma \sin\xi e^{i\varsigma t} \quad -i\delta \cos\xi e^{-i\delta t} \right)$$
$$\times \begin{pmatrix} e^{-i\varsigma t} \sin\xi \\ e^{i\delta t} \cos\xi \end{pmatrix} = i\varsigma (\sin x)^2 - i\delta (\cos x)^2 = 0 \tag{26}$$

The equivalent equations (25) and (26) lead to sufficient condition for parallel transport in the above example:

$$(\tan \xi)^2 = \frac{\delta}{\varsigma} \quad ; \quad \cos^2 \xi \left(1 + \frac{\delta^2}{\varsigma^2}\right) = 1 \rightarrow \cos \xi = \sqrt{\frac{\varsigma^2}{\delta^2 + \varsigma^2}} \quad . \quad (27)$$

Using Eq. (21) for $U(t)$ and Eq. (11) we get:

$$\langle 1|_0 U(t)|1\rangle_0 = \cos \xi e^{i(\delta t + \phi)} \quad ; \quad \langle 2|_0 U(t)|2\rangle_0 = \cos \xi e^{-i(\delta t + \phi)}$$

$$\gamma_{geom} = \arg\left[\sum_k w_k v_k e^{i\beta_k}\right] = \arg\left[w_1 \cos \xi e^{i(\delta t + \phi)} + w_2 \cos \xi e^{-i(\delta t + \phi)}\right] \quad . \quad (28)$$

The geometric phase is obtained by the incoherent superposition of two fields produced separately from the two components of the thermal field as given by Eq. (28), Such incoherent superposition but different system, was used in [9] for coupling it with Mach-Zehnder interferometer.

We evaluate the geometric phase of Eq. (28) which correspond to (12) as:

$$\gamma_{geom} = \arctan\left[\left(\frac{w_1 - w_2}{w_1 + w_2}\right) \tan(\delta t + \phi)\right] \quad (29)$$

and the field intensity is reduced by the factor:

$$\tilde{v} = \cos \xi \sqrt{w_1^2 + w_1^2 + 2 w_1 w_2 \cos(2\delta t + 2\phi)} \quad . \quad (30)$$

The unitary development of the SU(2) 1/2 system was produced by the consecutive interactions of this system with magnetic fields in times which are short relative to relaxation times in this system. Parallel transport equations, which eliminate the dynamical phase, are obtained by assuming the relation of Eq. Eq. (27) between the parameters included in the analysis.

## 4. Summary and Conclusion

The density matrix $\rho_0$ for mixed states given in Eq. (1) includes averaging of thermal fluctuations with probabilities $w_k$. Therefore, unitary transformations cannot operate directly on this density matrix, but they can operate separately on the components of such density matrix composed of pure quantum states. We reviewed the methods which are used for performing such transformations. The generalization of parallel transport equations to mixed states are of special interest as by their use dynamical phases are eliminated [9]. We find that the geometric phases obtained by the unitary transformation are Pancharatnam's phases which are gauge invariant [2] and are valid also for open circles. Our interest in the present work was in the two-dimensional thermal states. In addition to the interaction of a magnetic moment and constant magnetic field in the z direction which leads to an initial mixed state we include unitary interactions with magnetic fields interacting in short time relative to relaxation times. Using such system, we applied the unitary transformation given by Eq. (19). By applying the parallel transport equations, the interference intensity $\tilde{v}$ is reduced as given by Eq. (30) and the geometric phase is derived in Eq. (29), It is interesting to find that geometric phases can be obtained in systems which are different from the common ones.